\documentclass[twocolumn,twoside,letterpaper,11pt]{article}
\usepackage{graphicx,xrrc,natbib}
\usepackage{times}
\begin{document}

\newcommand{\la}{\,\rlap{\raise 0.5ex\hbox{$<$}}{\lower 1.0ex\hbox{$\sim$}}\,}
\newcommand{\ga}{\,\rlap{\raise 0.5ex\hbox{$>$}}{\lower 1.0ex\hbox{$\sim$}}\,}
\newcommand\degr{\hbox{$^\circ$}}

\title{PROBING COLLIDING WIND BINARIES WITH HIGH-RESOLUTION X-RAY SPECTRA}

\author{    D. B. Henley                                                  } 
\institute{ School of Physics \& Astronomy, The University of Birmingham  } 
\address{   Birmingham, B15 2TT, U.K.                                     } 
\email{     dbh@star.sr.bham.ac.uk                                        } 

\author{    I. R. Stevens, J. M. Pittard, M. F. Corcoran, A. M. T. Pollock}
\email{     irs@star.sr.bham.ac.uk, jmp@ast.leeds.ac.uk, 
            corcoran@barnegat.gsfc.nasa.gov, apollock@xmm.vilspa.esa.es}

\maketitle

\abstract{
X-ray line profiles represent a new way of studying the winds of massive stars.
In particular, they enable us to probe in detail the wind-wind collision in
colliding wind binaries, providing new insights into the structure and dynamics
of the X-ray-emitting regions. We present the key results of new
analyses of high-resolution \textit{Chandra} X-ray spectra
of two important colliding wind systems, $\gamma^2$~Velorum and WR140.
The lines of $\gamma^2$~Vel are essentially unshifted from their rest wavelengths,
which we suggest is evidence of a wide shock opening angle, indicative of sudden
radiative braking. The widths of the lines of WR140 are correlated with ionization
potential, implying non-equilibrium ionization. The implications of these results
for the radio emission from these systems are discussed, as are some of the future
directions for X-ray line profile modelling of colliding wind binaries.
}


\section{Introduction}

Massive, early-type stars ($M \ga 10M_\odot$) have profound effects on their
environments through their intense UV radiation fields, their powerful stellar
winds ($\dot{M} \sim 10^{-7}$--$10^{-5}~M_\odot~\mathrm{yr}^{-1}$,
$v_\infty \sim 2000~\mathrm{km~s}^{-1}$) and their violent deaths as supernovae.
The winds from massive stars provide $\sim$30\% of the mechanical energy input into
the interstellar medium \citep{abbott82}, and the mass-loss also substantially affects
the stars' own evolution \citep{chiosi86}. Thus, understanding mass-loss from
massive stars is essential for understanding feedback to the ISM, stellar evolution
and galactic evolution.

Most massive stars reside (or resided) in binaries or multiple systems \citep{zinnecker03}.
In such systems, the winds of the massive stars collide highly supersonically
(Mach number $\ga$ 100), and they are compressed and shock-heated to $\sim$$10^7$~K.
The shock-heated gas produces copious X-rays \citep*[e.g.][]{stevens92}, though
there are observational signatures throughout the electromagnetic spectrum,
including non-thermal radio emission from relativistic electrons accelerated at the
shocks \citep[e.g.][]{eichler93,dougherty00}.

Previously, studies of X-rays from colliding wind binaries (CWBs) have tended
to focus on their broad-band spectral properties, due to the poor spectral resolution
of the satellites then in operation (e.g. $E/\Delta E \sim 20$ for the \textit{ASCA}
SIS).  Nevertheless, by comparing the variable \textit{ASCA} spectrum of $\gamma^2$ Velorum
with hydrodynamical models of the system, \citet{stevens96} were able place constraints
on the stars' mass-loss rates and wind velocities. This technique has more recently been
applied to the high-resolution \textit{Chandra} grating spectrum of $\eta$ Carinae
\citep{pittard02a}.

\begin{figure*}
\includegraphics[width=18cm]{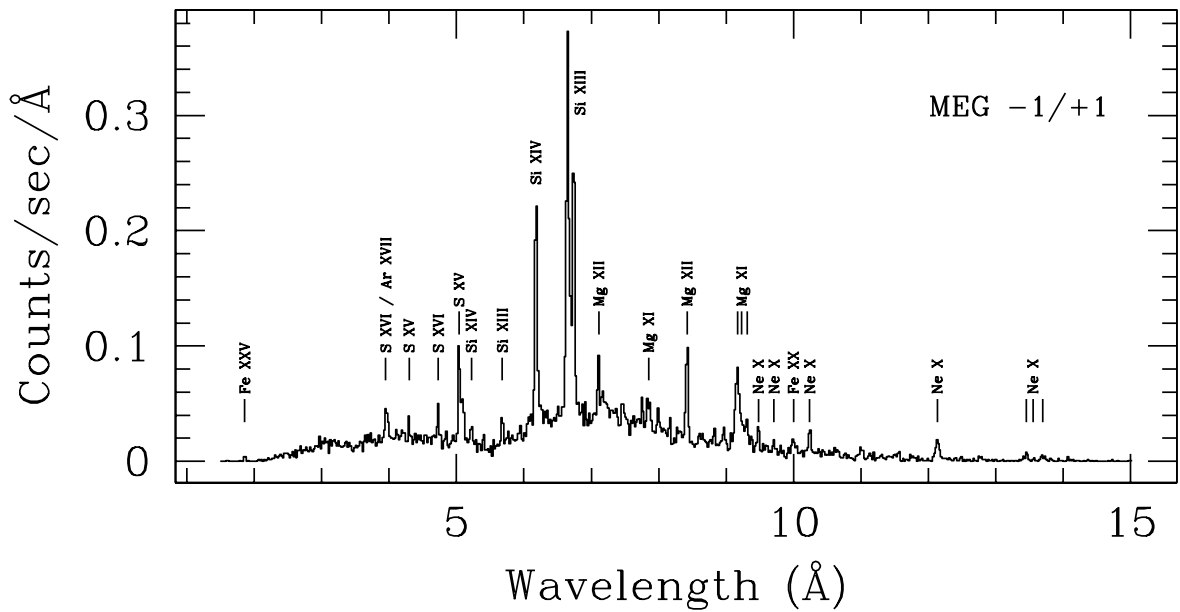}
\caption{The co-added first order MEG spectrum of $\gamma^2$~Vel. The spectrum has
been binned up to 0.02~\AA. The absorption at long wavelengths is due to the unshocked
stellar winds.}
\label{fig:gammaVelspectrum}
\end{figure*}

The unprecedented spectral resolution offered by the gratings on
board \textit{Chandra} and \textit{XMM-Newton} enables us to resolve line shifts and
widths down to a few hundred km~s$^{-1}$, providing a unique probe of the dynamics of
the wind-wind colllision. Furthermore, the X-ray forbidden-intercombination-resonance (\textit{fir})
triplets from He-like ions provide diagnostics of the density and temperature of the
X-ray-emitting plasma, as well as potentially providing diagnostics of
the UV radiation field \citep[e.g.][]{paerels03}.
Using these tools we can obtain new insights into the location, geometry, structure and
dynamics of the wind-wind collision region.


\section{$\gamma^2$ Velorum}
\label{sec:gammaVel}

We have carried out a new analysis of an archived 65-ks \textit{Chandra} High-Energy
Transmission Grating Spectrometer (HETGS) observation of the well-studied WR+O binary
$\gamma^2$~Velorum  (\citealp{skinner01}; Henley et al., in prep.). $\gamma^2$~Vel
is a double-lined spectroscopic binary of spectral type WC8 + O7.5 \citep{demarco99}
whose orbit is well determined, with a period of $78.53 \pm 0.01$~days, $e = 0.326 \pm 0.01$,
$\omega_\mathrm{WR} = 68\degr \pm 4\degr$ \citep{schmutz97} and $i = 63\degr \pm 8\degr$
\citep{demarco99}.

The first order Medium Energy Grating (MEG) spectrum of $\gamma^2$~Vel is shown in
Fig.~\ref{fig:gammaVelspectrum}. The spectrum is dominated by strong emission lines
from S ($\lambda \approx 4$--5~\AA), Si ($\lambda \approx 5$--7~\AA) and Mg
($\lambda \approx 7$--9~\AA), with weaker lines from Ne ($\lambda \approx 9$--14~\AA)
and Fe (e.g. $\lambda \approx 2$~\AA\ and $\lambda \approx 10$~\AA).

\begin{figure}
\includegraphics[width=8cm]{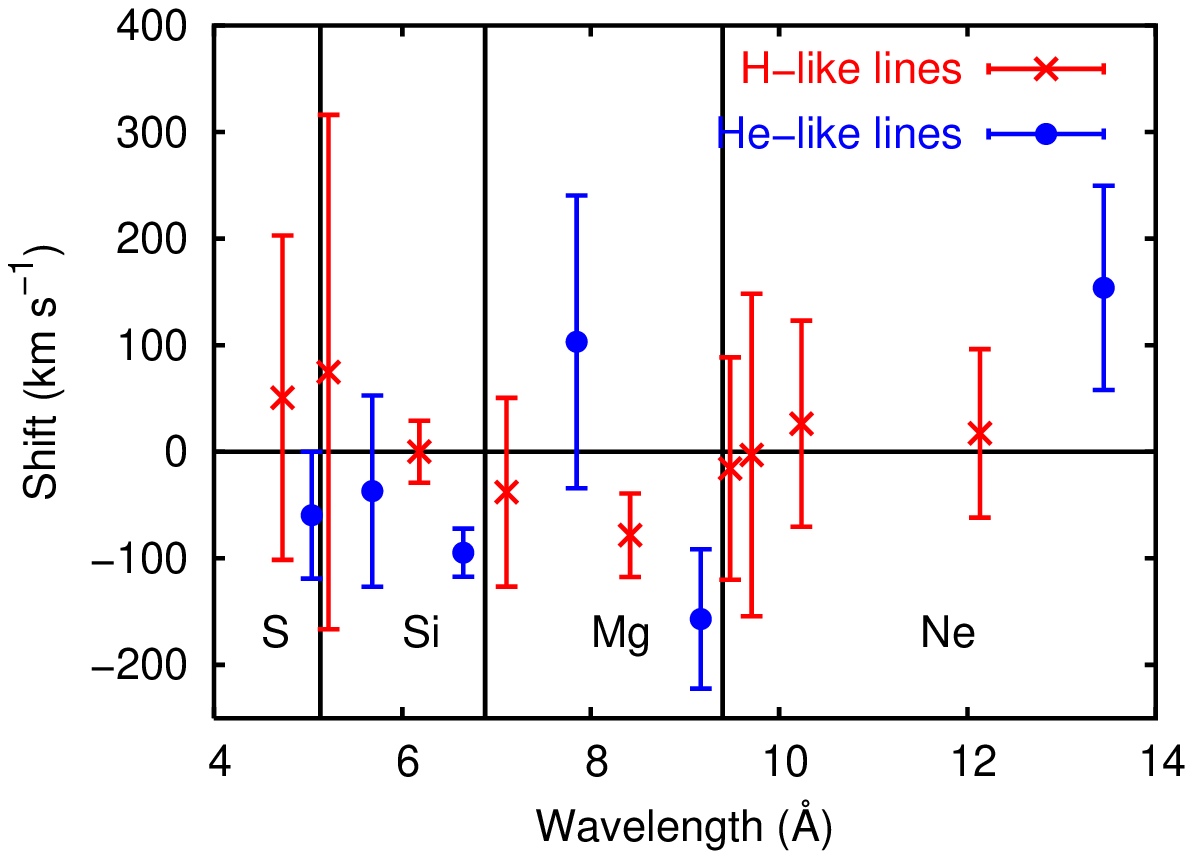}
\includegraphics[width=8cm]{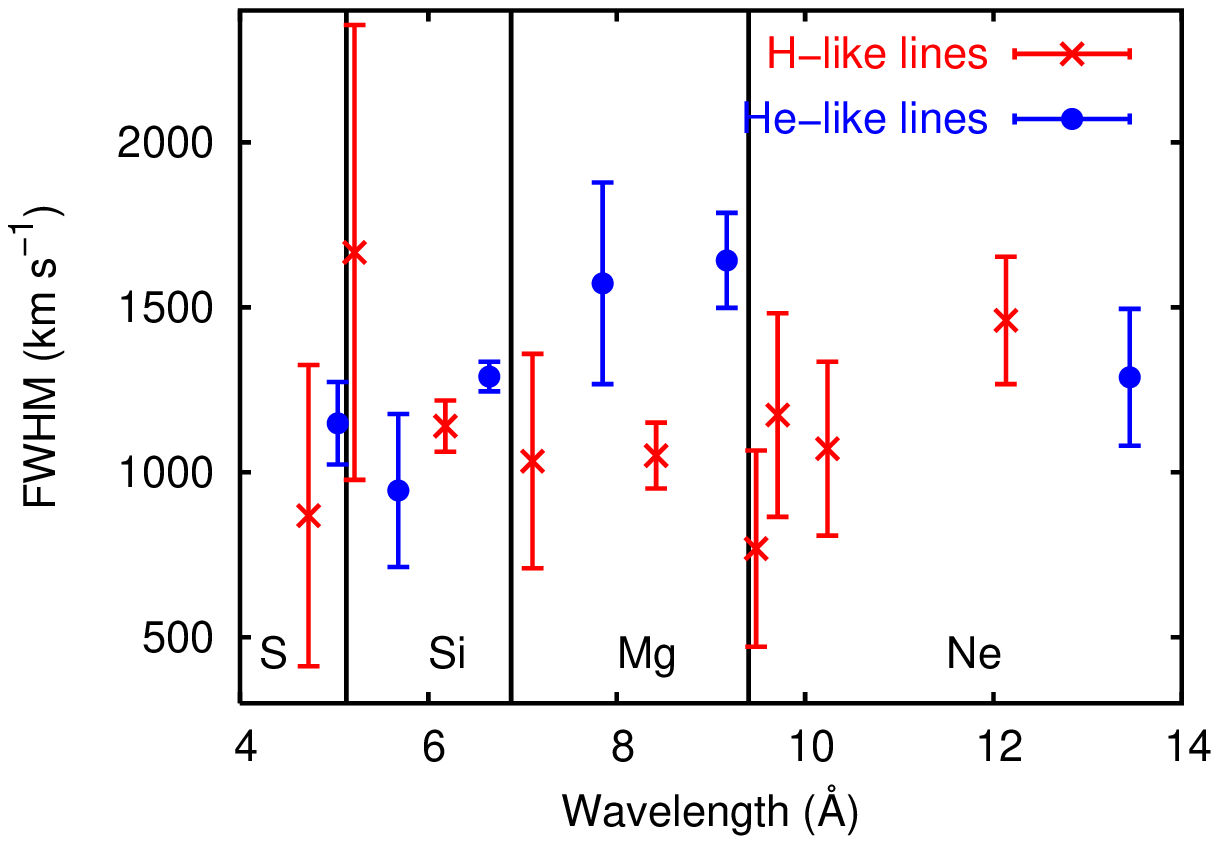}
\caption{The shifts and widths of the emission lines in the \textit{Chandra}-HETGS
spectrum of $\gamma^2$~Vel, plotted as a function of lab wavelength.}
\label{fig:gammaVelshiftsandwidths}
\end{figure}

We have measured the line centroid shifts and the line widths by fitting Gaussian line
profiles (plus a constant continuum) to a narrow wavelength region around the particular
line of interest. Although colliding wind binaries are expected to exhibit a wide range of
line profile shapes \citep*{henley03}, Gaussians give good fits to the emission lines in
$\gamma^2$~Vel and thus provide a good way of quantifying the line shifts and widths.

The measured line shifts and widths are plotted against wavelength
in Fig.~\ref{fig:gammaVelshiftsandwidths}.
The lines are generally unshifted from their lab wavelengths. The mean shift is
$-52 \pm 16 \mathrm{~km~s}^{-1}$ and the mean FWHM is $1240 \pm 40 \mathrm{~km~s}^{-1}$.
Neither the line shift nor width is correlated with wavelength or ionization potential.

\begin{figure}
\includegraphics[width=8cm]{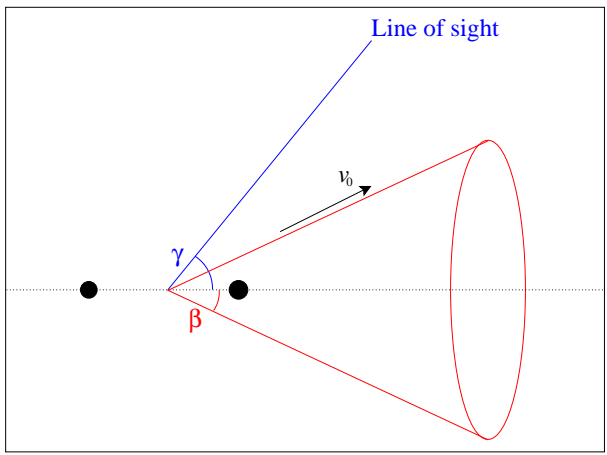}
\caption{A simple geometrical model for the wind-wind collision in a CWB. The black
circles denote the two stars and the red cone (with opening half-angle $\beta$) is
the wind-wind interaction region (along which X-ray-emitting material is streaming
at speed $v_0$).}
\label{fig:geometricalmodel}
\end{figure}

In order to understand our measured shifts and widths in terms of the geometry of the
wind-wind collision, we use a simple model in which the wind-wind interaction region
is a conical surface \citep{luhrs97} with opening half-angle $\beta$. The X-ray-emitting
gas streams along this cone away from the line of centres at speed $v_0$
(see Fig.~\ref{fig:geometricalmodel}). Assuming the material is distributed symmetrically
around the shock cone, the line centroid shift ($\bar{v}$) and width are given by
\citep{pollock04}
\begin{equation}
\hspace{14.5mm} \bar{v} = - v_0 \cos \beta \cos \gamma
\end{equation}
\begin{equation}
\mathrm{FWHM} = v_0 \sin \beta \sin \gamma
\end{equation}
The angle $\gamma$ can be calculated from the orbit \citep{schmutz97,demarco99};
$\gamma = 44\degr$ at the time of the \textit{Chandra} observation. By eliminating
$v_0$ from the above equations, and using our measured shifts and widths
($\bar{v} = -52 \mathrm{~km~s}^{-1}$, $\mathrm{FWHM} = 1240 \mathrm{~km~s}^{-1}$)
we find that $\beta = 88\degr$. Note, however, that the mean observed line shift is
comparable to the absolute wavelength accuracy of the HETGS
($\sim$100~km~s$^{-1}$; \textit{Chandra} Proposers' Observatory Guide). This means that
the measured value of $\bar{v}$ (and hence the derived value of $\beta$) may not be completely
trustworthy. However, we can say that $| \bar{v} | < 100 \mathrm{~km~s}^{-1}$ (since we would
unambiguously be able to detect a larger value of $\bar{v}$). This implies that $\beta > 85\degr$.

\begin{figure}
\includegraphics[width=8cm]{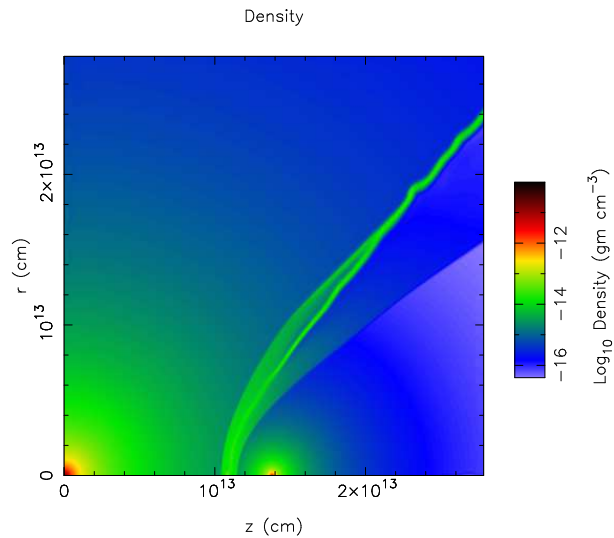}
\caption{Density map from a hydrodynamical simulation of $\gamma^2$~Vel.
The adopted wind parameters are
$\dot{M}_\mathrm{WR} = 1 \times 10^{-5} ~M_\odot~\mathrm{yr}^{-1}$,
$\dot{M}_\mathrm{O} = 5 \times 10^{-7} ~M_\odot~\mathrm{yr}^{-1}$,
$v_\mathrm{WR} = v_\mathrm{O} = 1500 ~\mathrm{km~s}^{-1}$.}
\label{fig:gammaVeldensity}
\end{figure}

To understand this value in terms of the dynamics of
the system, we have carried out a hydrodynamical simulation of the wind-wind
collision. For simplicity we have assumed non-accelerating, spherically symmetric
winds. Since there is
evidence that the O star wind collides below its terminal velocity
\citep*[e.g.][]{stlouis93}, we have adopted an O star wind speed of 1500~km~s$^{-1}$
(cf. $v_\infty  \approx 2300$--2400~km~s$^{-1}$; \citealp*{prinja90}; \citealp{stlouis93}).
We assume the WR star collides at its terminal velocity \citep*[1500~km~s$^{-1}$;][]{barlow88}.
We adopt mass-loss rates of $\dot{M}_\mathrm{WR} = 1 \times 10^{-5} ~M_\odot~\mathrm{yr}^{-1}$
and $\dot{M}_\mathrm{O} = 5 \times 10^{-7} ~M_\odot~\mathrm{yr}^{-1}$.
Figure~\ref{fig:gammaVeldensity} shows a density map from the resulting simulation.

We have used this simulation to model the X-ray line emission from $\gamma^2$~Vel
using the model described in \citet{henley03}, which calculates X-ray emission line
profiles from the hydrodynamical simulation results assuming collisional ionization
equilibrium.  The resulting profiles are typically blueshifted by $\sim$300~km~s$^{-1}$,
in contrast to the essentially unshifted lines in the observed spectrum.

The discrepancy between the observed and calculated line shifts is most likely due to
the shock opening half-angle in the hydrodynamical simulation being $\sim$40\degr, whereas the
simple geometrical model described above implies a shock opening half-angle of
$>$85\degr. In a CWB with non-accelerating winds, $\beta$ is a function only of the wind
momentum ratio ($\dot{M}_\mathrm{WR} v_\mathrm{WR} / \dot{M}_\mathrm{O} v_\mathrm{O}$).
As this ratio increases, $\beta$ decreases, because the more powerful wind of the WR star
overwhelms that of the O star. The shock opening angle in our hydrodynamical simulation
therefore depends on the adopted values of the wind parameters. However, the shock opening
half-angle inferred from the simple geometrical model ($\beta > 85\degr$) implies approximately
equal wind momenta, which is not the case for any sensible set of (constant velocity) wind
parameters. We therefore suggest that this large implied opening
angle may be evidence of sudden radiative braking \citep*{owocki95,gayley97},
in which the wind of the WR star is rapidly decelerated when it encounters the radiation
field of the O star. This rapid deceleration leads to an increase in the shock opening angle
(the larger the opacity of the WR star wind to the O star radiation field, the larger
the shock opening angle; see Fig.~3 in \citealp{gayley97}). It also alters the Mach number
of the wind-wind collision, resulting in softer X-ray emission than would otherwise be
expected. However, as well as affecting the X-ray emission, radiative braking may also
affect the non-thermal radio emission. This is because the orbital variability of the
non-thermal radio emission (due to free-free absorption in the stars' unshocked winds)
is dependent on the shock opening angle.

\begin{figure*}
\includegraphics[width=18cm]{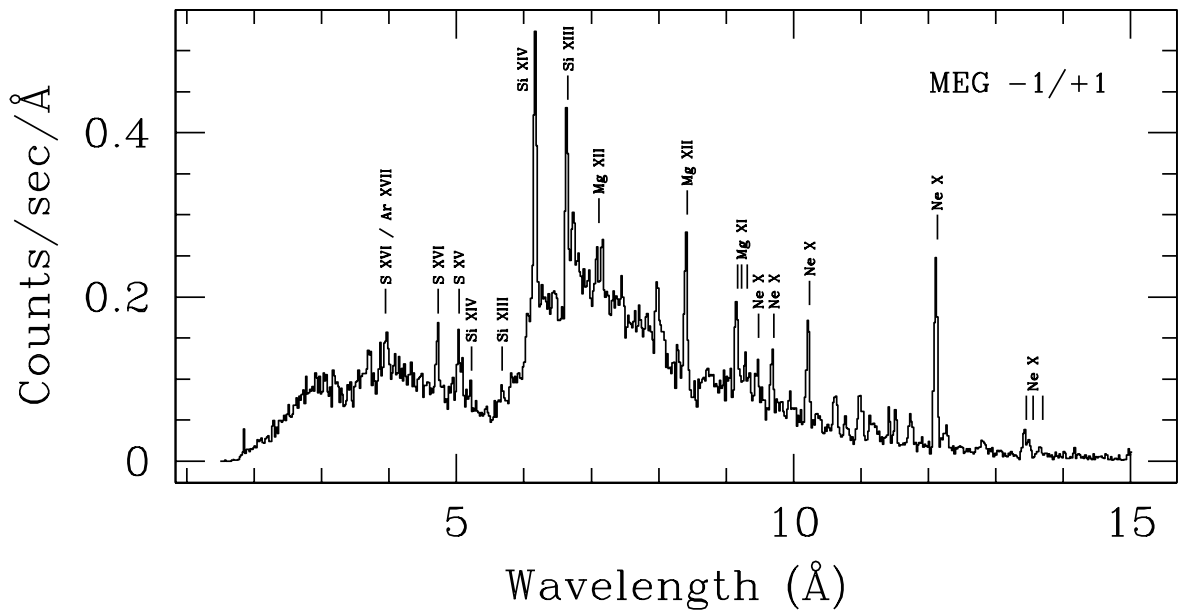}
\caption{The co-added first order MEG spectrum of WR140. The spectrum has
been binned up to 0.02~\AA. The absorption at long wavelengths is due to the ISM.}
\label{fig:wr140spectrum}
\end{figure*}

\citet{chapman99} analysed ATCA radio observations of $\gamma^2$~Vel at four wavelengths using
a thermal$~+~$non-thermal emission model. The thermal free-free emission is from the unshocked
stellar winds, while the non-thermal synchrotron emission is from the wind-wind collision and
is attenuated by free-free absorption in the unshocked winds. \citet{chapman99} found that
over the range $\sim$2--4~GHz, thermal and non-thermal emission make approximately equal
contributions to the observed flux. However, their observations were made at two different
orbital phases, and so their assumptions of constant non-thermal spectral index and
constant optical depth for the non-thermal emission may not be valid.

Radio observations with better orbital coverage (S. Dougherty, priv. comm.) show no evidence of
non-thermal emission from $\gamma^2$~Vel. This is presumably because $\gamma^2$~Vel is a
relatively close binary, and so the non-thermal emission from the wind-wind collision is
strongly absorbed by the unshocked stellar winds. Also, the population of shock accelerated
relativistic electrons will be quickly reduced by efficient inverse Compton cooling.
Thus, it may unfortunately be the case
that the effects of radiative braking on the non-thermal radio emission discussed
above will be very difficult to observe, because radiative braking is favoured in
close systems \citep{gayley97} in which any non-thermal radio emission from the wind-wind
collision will be strongly attenuated by the unshocked winds.


\section{WR140}

The eccentric WC7 + O4.5 binary WR140 ($P = 2900 \pm 10$~days, $e = 0.84 \pm 0.04$;
\citealp{williams90}) is another key system in our understanding of CWBs.
Compared with $\gamma^2$~Vel, WR140 is hotter and about two orders of magnitude more luminous.
\textit{ROSAT} and \textit{ASCA} observations \citep{corcoran96b}, and more
recently \textit{RXTE} observations \citep{pollock04} show a gradual rise in X-ray
flux from apastron to periastron, followed by a rapid drop in flux at
around the time of periastron passage. The flux variations
agree qualitatively with the theoretical predictions of \citet{stevens92} -- the
increase in flux prior to periastron is due to an increase in the density of the
X-ray-emitting gas as the stars approach each other, whereas the subsequent sudden
drop in flux is due to increased absorption as the line of sight moves from the O
star wind into the denser WR star wind.

\begin{figure}
\includegraphics[width=8cm]{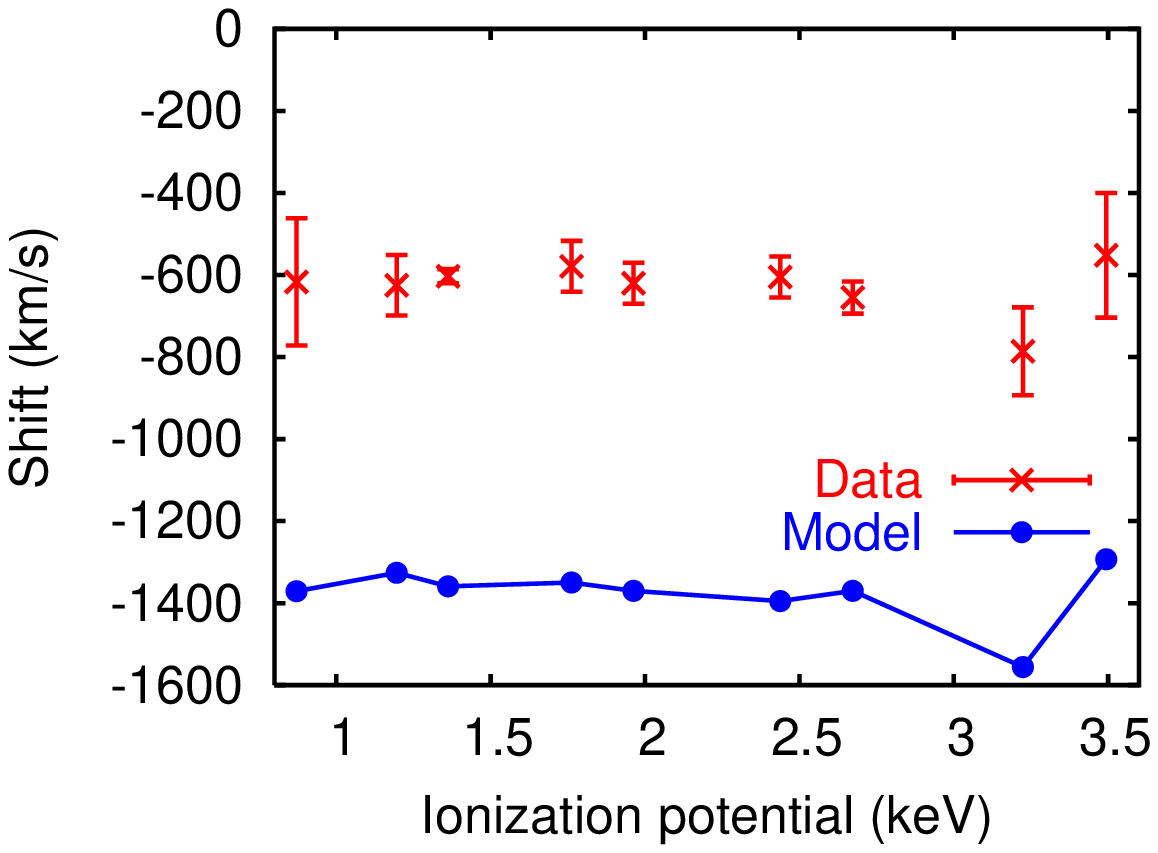}
\includegraphics[width=8cm]{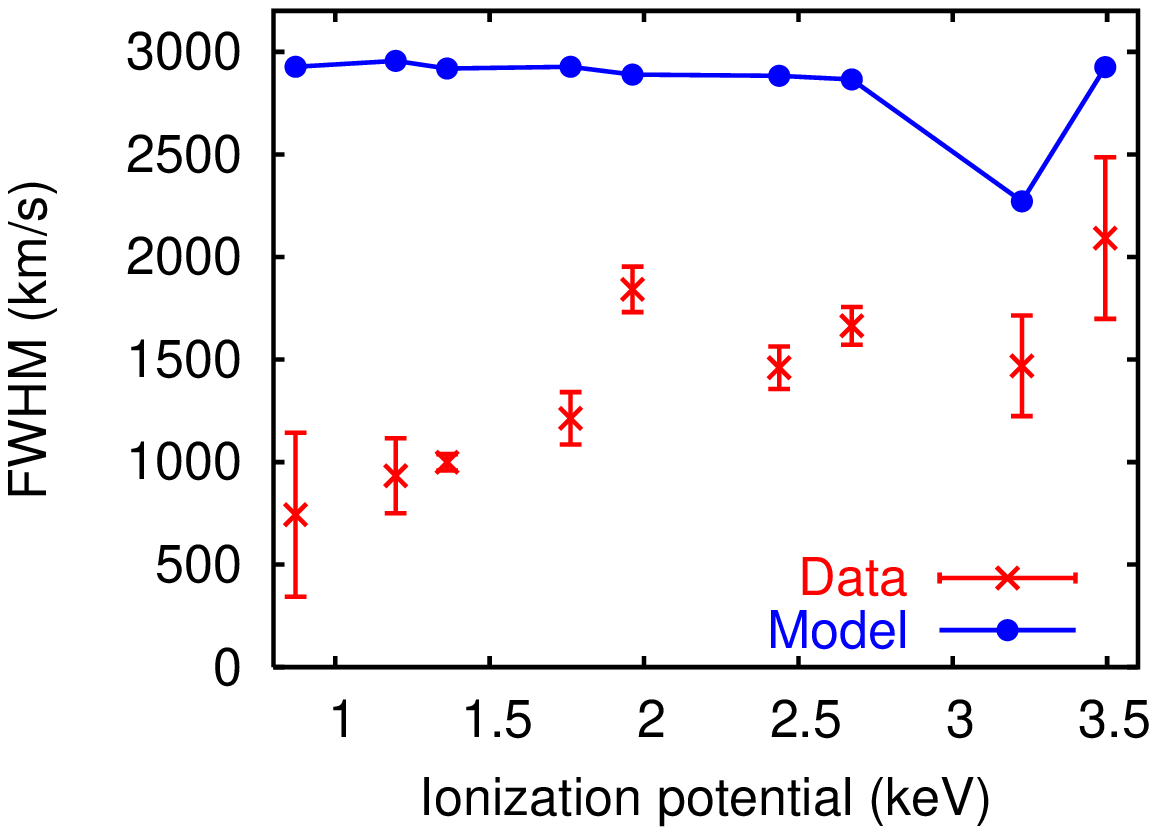}
\caption{The shifts and widths of the emission lines in the \textit{Chandra}-HETGS
spectrum of WR140 compared with shifts and widths calculated using the model of
\citet{henley03}.}
\label{fig:WR140shiftsandwidths}
\end{figure}

Figure~\ref{fig:wr140spectrum} shows the first-order MEG spectrum from a 45-ks
\textit{Chandra}-HETGS observation of WR140 obtained a few weeks before its most recent
periastron passage \citep{pollock04}. As for $\gamma^2$~Vel, line shifts and widths
were measured by fitting Gaussians to individual emission lines. The measured shifts
and widths are shown in Fig.~\ref{fig:WR140shiftsandwidths}, plotted against ionization
potential. The lines are systematically blue-shifted
by $\sim$600~km~s$^{-1}$, and the line widths are correlated with ionization potential.

As for $\gamma^2$~Vel, we have attempted to model the X-ray line emission from WR140 using 
the model of \citet{henley03}, again assuming spherically symmetric winds and collisional
ionization equilibrium. The wind parameters used are from \citet{stevens92}.
Radiative braking is not an issue in WR140, as it is a
much wider binary than $\gamma^2$~Vel [at the times of their respective \textit{Chandra}
observations, the stellar separations were 0.9~A.U. ($\gamma^2$~Vel) and 4.2~A.U.
(WR140)].  The calculated line shifts and widths are
compared with the observed values in Fig.~\ref{fig:WR140shiftsandwidths}.
As can be seen, there is poor agreement between the predictions of the model and the
observed shifts and widths. Firstly, the observed shifts are smaller than expected.
This implies that the velocity along the shock cone is lower than expected. Indeed,
the highest velocity observed in the resolved lines ($\sim$2000~km~s$^{-1}$) is
significantly lower than the velocities of the unshocked winds
\citep[$\sim$3000~km~s$^{-1}$;][]{pollock04}.
This may mean that some of the energy budget is going into the
production of relativistic particles. Secondly, the correlation between the line widths
and ionization potential is not reproduced by the model. This may be indicative of
non-equilibrium ionization, as it implies that higher excitation lines (e.g. S~XVI)
originate further from the line of centres where the velocity is larger. This is
contrary to what is expected if collisional ionization equilibrium holds, because
the temperature of the shocked gas in the wind-wind collision decreases monotonically away
from the line of centres (hence lines from more highly ionized species would be expected
to originate nearer the line of centres).

Over the course of eight years at the VLA, \citet{white95} obtained a detailed radio
light curve of WR140. The observed flux is non-thermal, and is highly variable at all
three wavelengths observed (2, 6 and 20~cm). The radio emission peaks before periastron
passage, at an orbital phase of $\phi \sim 0.7$--0.8 (where $\phi = 0$ corresponds to
periastron). The 2~cm emission peaks earliest, and is already on the decline while
the 20~cm emission is still rising.

\citet{white95} were unable to explain the variation in
the radio flux (which they attributed to both a varying intrinsic luminosity and
a varying optical depth) with spherically symmetric winds. Instead they proposed a model in
which the wind of the WR star is disk-like. The shifts and widths of the X-ray emission
lines from WR140 are also not well described by our current model with spherically symmetric
winds, as discussed above.
This may lead one to speculate whether or not they can also be explained by a disk-like
wind model. However, \citet{white95} essentially assume the non-thermal source
is point-like, whereas theoretical modelling of the radio emission from wind-wind collisions
shows that it is important to consider extended emission regions \citep{dougherty03}.
Therefore, it may be possible to explain WR140's radio lightcurve with spherically
symmetric winds, without having to invoke a disk-like wind model.
Furthermore, the variation in the
observed X-ray flux is in good qualitative agreement with the predictions of a spherically
symmetric colliding wind model \citep{corcoran96b,pollock99}.
It seems that more detailed modelling of the X-ray emission lines and the radio emission,
and more spectral information from the radio are required to resolve the
issue of the nature of WR140's winds. (See also Pittard et al., these proceedings.)


\section{Conclusions}

High-resolution X-ray spectra allow us to probe the structure and dynamics of the
wind-wind collision in CWBs. In the two key systems described here, the X-ray spectra have
revealed evidence of a variety of interesting phenomena, such as sudden radiative braking and
non-equilibrium ionization. From a radio point-of-view, the former is of particular
interest, as radiative braking increases the shock opening angle, which in turn
influences the variability of the non-thermal emission.

Future modelling of the X-ray emission lines needs to take into account the effects
of the stars' radiation fields and the acceleration of relativistic particles on the
hydrodynamics, as well as calculating the ionization balance self consistently (rather
than assuming collisional ionization equilibrium). Furthermore, while line emission is
generally considered a thermal process, ions may equally well be excited by collisions
with non-thermal electrons. If the relative abundance of non-thermal electrons is
large (the energy density of relativistic electrons in CWBs may be as large as $\sim$1\%\ of the
thermal energy density; see \citealp{dougherty03}) these too will have to be included
in the models.

\section*{Acknowledgments}

DBH gratefully acknowledges funding from the School of Physics \& Astronomy
at the University of Birmingham.

\end{document}